\documentclass[12pt]{article}
\usepackage{amsmath,amsfonts, amssymb,graphicx,geometry,authblk,color,soul,comment,hyperref,wrapfig} 
\title{Multiscale modeling of materials and neural operators}
\author{Kaushik Bhattacharya\footnote{Corresponding author: bhatta@caltech.edu}}
\affil{California Institute of Technology, Pasadena, CA 91011, USA}

\begin{document}
\maketitle
\begin{abstract}
Multiscale modeling is essential for understanding the complex behavior of materials.  However, accurately transferring all relevant information from one scale to another has remained an outstanding challenge.  Neural operators, discretization-independent generalizations of neural networks, is proving to be a powerful tool in addressing this challenge.  This article provides an introduction to neural operators, and illustrates their use in multiscale modeling of materials through three selected examples.

\end{abstract}


\section{Introduction}
The properties of materials that we observe are the result of phenomena at various length and time scales starting from quantum mechanics and going through atoms, defects and microstructure \cite{phillips_2001}.  Consider, for example, the strength of metallic alloys.  Plasticity is mediated by dislocations.  Quantum mechanical details play a critical role in determining the structure of the core, while the elastic fields they generate can interact with macroscopic strain fields.  Consequently dislocations interact with solutes, with precipitates, with grain boundaries and each other.  Large ensembles of dislocations give rise to plasticity at the level of a single grain, but the grains act collectively as allowed by the crystallographic texture to determine the strength of the metal.  

Historically, we have built models at each scale to understand specific mechanisms.   For example, different aspects of dislocations have been studied with first principles theories \cite{woodward_2008,das_2019}, atomistic theories \cite{maresca_2024}, and as lines in an elastic medium.  The collective behavior of ensembles has been studied with dislocation dynamics \cite{giessen_1995,bulatov_2013} and field theories \cite{berdichevskii_1967,le_2016}, and the behavior of polycrystals with crystal plasticity \cite{asaro_1983,lebensohn_2020}.  However, these theories focus on one specific scale and mechanism, using its own specific language, mathematical models and computational methods.

Over the last three decades, multiscale modeling has tried to understand the relationship between these scales and mechanisms \cite{fish_2009,van_2020,radhakrishnan_2021}.  It creates a linear order of mechanisms from the fine to the coarse scale, and then postulates that mechanisms only interact pairwise.  Further, the coarser scale regulates the finer scale by providing the overall conditions, and then filters the detailed information from the finer scale to obtain the behavior of the finer scale that is relevant to the coarser scale.  While there are exceptions \cite{tadmor_1996,car_1985}, these methods typically assume a separation of scales.  Statistical mechanics, homogenization, coarse graining and other theories provide the  framework for such a procedure \cite{pavliotis_2008,tadmor_2011,e_2011}.  Though many mathematical questions remain open, the framework has proved enormously useful in understanding material behavior and in the accelerated development of materials.

Still, the systematic implementation of multiscale modeling remains a challenge.  As formulated, the fine scale model has to be evaluated at each state of the coarse model.  This proves prohibitively expensive.  Therefore, the typical practice is to build an {\it ad hoc} empirical model at the coarse scale and then use the small scale to fit the parameters of this coarse scale model.  Unfortunately, this practice runs counter to the purpose of multiscale modeling which is to incorporate finer scale information with no outside information.  Worse, there is no mathematical notion of accuracy or convergence in such a procedure, and thus this approach is limited by model or epistemic uncertainty.  

This article discusses an alternate approach, one that has emerged in recent years, and seeks to use machine learning in multiscale modeling \cite{lefik_2009,radhakrishnan_2021,peng_2021,ingolfsson_2023}.  However,  there are difficulties in using standard neural networks that were motivated by image processing and language.  Instead, we show that neural operators (generalization of deep neural networks to function spaces) provide a highly efficient and accurate approach to transfer information from one scale to another.  Thus, neural operators enable a systematic approach towards high fidelity multiscale models.  The literature is vast and rapidly expanding, and therefore this is not intended as a comprehensive review.  Instead, we highlight the issues and principles with three examples selected from the work of the author and his collaborators.

\section{Multiscale modeling and surrogates}

\begin{figure}
\centering
\includegraphics[width=6in]{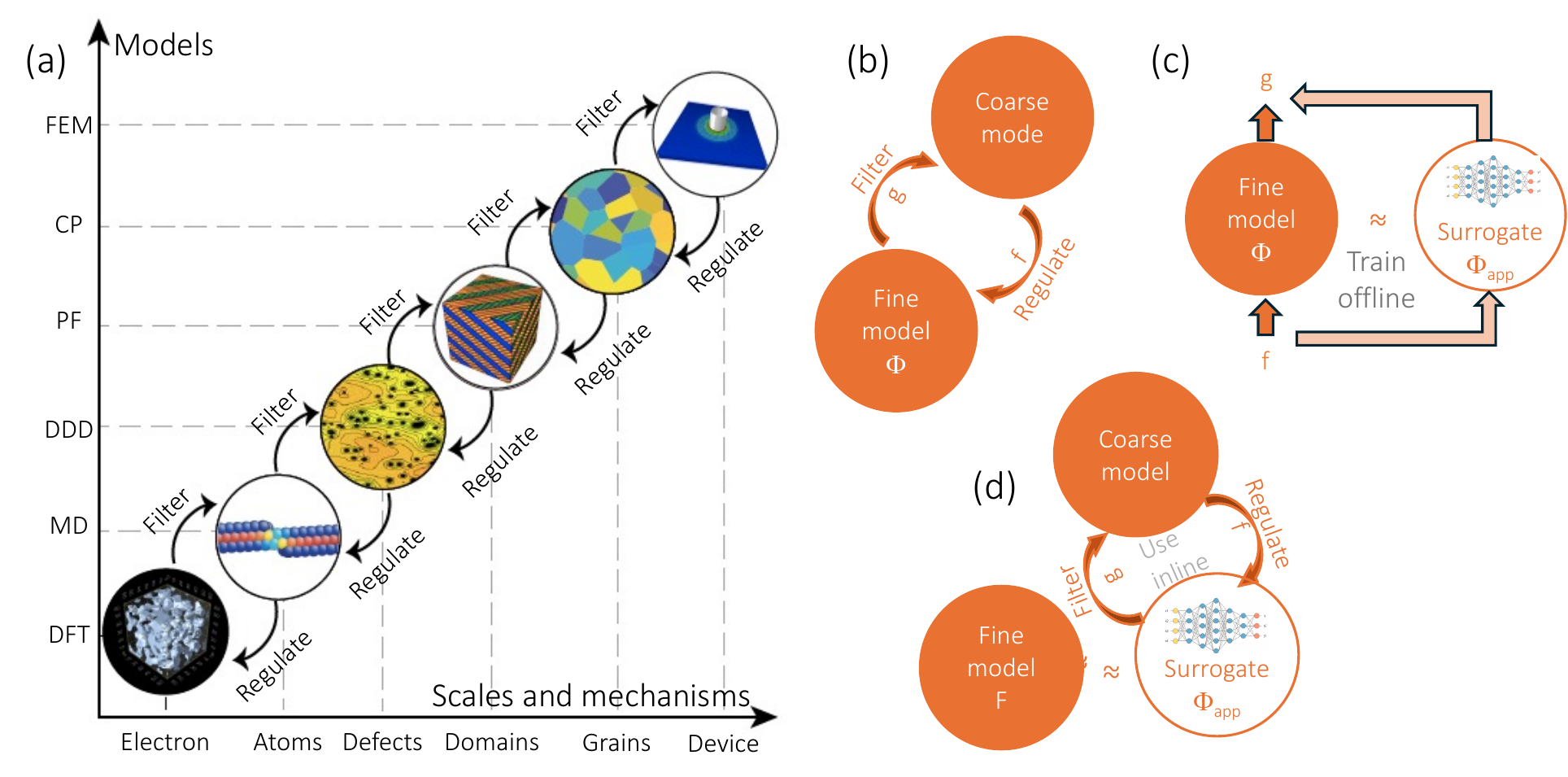}
\caption{\small Multiscale modeling and proposed approach.  (a) Multiscale model of strength (adapted from \cite{kovachki_2022}).  DFT=Density functional theory, MD=molecular dynamics, DDD = discrete dislocation dynamics, PF=phase field, CP=crystal plasticity, FEM = Finite element method.  (b) Canonical problem of multiscale modeling.  (c) The computationally expensive fine scale model is replaced with an inexpensive but accurate surrogate.  (d) The surrogate replaces the fine scale model in  multiscale calculations.  \label{fig:approach}}
\end{figure}

A multiscale model of strength is shown in Figure \ref{fig:approach}(a) \cite{kovachki_2022}.  The strength of an alloy is the result of a hierarchy of mechanisms at distinct length and time scales.  We postulate that only two scales interact at a time.  This leads to the canonical problem shown in Figure \ref{fig:approach}(b).  We have a coarse model and a fine model.  The coarse model regulates the fine model (for example by providing boundary conditions or average values), we solve the fine model $\Phi$ at  high fidelity, and then filter the results (for example by averaging) and return it to the coarse model, thereby closing the two scale model.  This is then applied throughout the hierarchy.  The difficulty is that we have to do this at each computational point of the coarse model at each time step.  This is not feasible, even with the massive explosion of computational ability, especially if one seeks to maintain necessary numerical accuracy.  

So, we seek to replace the small scale model with a surrogate model $\Phi_\text{app}$, that is accurate but inexpensive to evaluate, Figure \ref{fig:approach}(c).  Classically, this surrogate has been a phenomenological model with a few adjustable parameters.  However, such models are limited by imagination and preconceptions, and ultimately defeat the whole multiscale modeling philosophy of incorporating fine scale physics without bias or prejudice.  We therefore seek a data-driven approximation.  We evaluate the fine model $\Phi$ over a large number $N$ of inputs $\{f_i\}_{i=1}^N$, and we use the resulting input-output pairs $\{f_i, g_i\}_{i=1}^N$ as data to train an accurate approximation  $\Phi_\text{app}$ that is inexpensive to evaluate.  If we can successfully do so, we can then use $\Phi_\text{app}$ as a surrogate in the coarse scale simulations as shown in Figure \ref{fig:approach}(d).

The key challenge, then, is to find a form for the approximation $\Phi_\text{app}$ that is sufficiently rich to provide an unbiased approximation, that can be trained to the desired level of accuracy using data, and that is sufficiently inexpensive to evaluate on the fly so that it can be used in multiscale simulations.  Neural networks have been recognized to have these properties, and it is natural to use them as candidates for $\Phi_\text{app}$.

Unfortunately there is a subtle problem that makes neural networks unsuitable as a potential surrogate.  A neural networks maps a finite dimensional vector to a finite dimensional vector.  However, our input $f$ as well as our output $g$ can be functions.  Consider for example the largest scale of crystal plasticity in Figure \ref{fig:approach}(a).  The input to the fine model is average strain history (since plasticity is history dependent), and the output is  the average stress history -- these are functions of time.  Now, in computation, we use time discretization, and therefore, both the input and the output are finite dimensional vectors.  So we can train a neural network.  Unfortunately, being an unbiased approximation, the neural network not only learns the response but also the time discretization.  Therefore, we can only use it at that time discretization: it has significant, uncontrolled errors if used at a different discretization.   This is a devastating problem in multiscale modeling: it is often necessary to use a fine time discretization to resolve all the fine details of the fine model, but use large time steps in the coarse model to be able to span the time intervals necessary in application.  Thus, we need {\it our surrogate to be accurate independent of discretization.  Unfortunately neural networks fail} in this requirement.  Neural operators are generalization of neural networks that map functions to functions, and are discretization independent.

\section{History dependent behavior in crystal plasticity and recurrent neural operator}

\begin{figure}[t]
\centering
\includegraphics[width=6in]{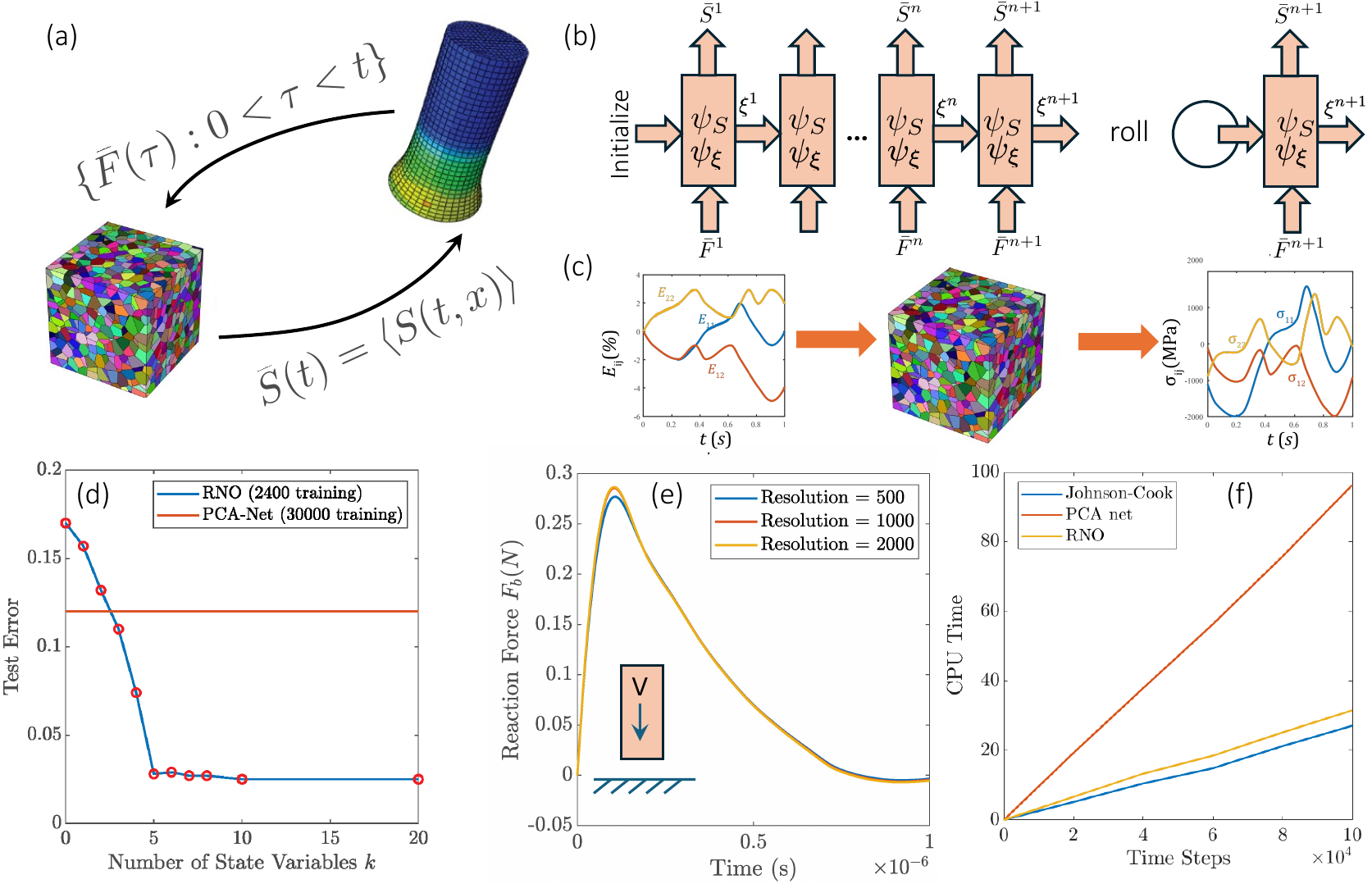}
\caption{\small Metal plasticity.  (a) The two scale problem.  A typical metal is made of a large number of grains.  The coarse model is the continuum model and the fine model is a polycrystal made of a large number of grains.  (b) Recurrent neural operator (RNO) discretized in time.  (c) Generating data from crystal plasticity.  (d) Test error vs. number of state variables in the RNO.  (e) Force vs. time in an impact problem computed using the RNO as surrogate at various time discretization. (e) Computational performance of finite elements using an RNO compared to other methods.  Adapted from \cite{liu_2022,liu_2023}.  \label{fig:cp}}
\end{figure}

\subsection{Multiscale problem of crystal plasticity}

We begin with the problem of modeling polycrystalline magnesium subject to complex loading studied in Liu {\it et al.} \cite{liu_2022, liu_2023}.  The grains are much smaller than the specimen.   Magnesium has a hexagonal closed packed crystal structure, and the inelastic behavior of each grain is controlled by a number of slip and twinning modes.   However, since the grains are oriented differently from each other, one has a very complex field of elastic and inelastic deformation.  A full resolution computation would require us to resolve details at a scale smaller than that of a grain, but over a region that spans a large number of grains.  This is computationally infeasible, even with the largest computers.  However, we can take advantage of the separation of grain and specimen scales, and use the two-scale expansion.  We refer the reader to \cite{bensoussan_1978,pavliotis_2008} for a general introduction to the methods, and \cite{liu_2022} for details in this specific problem.  This expansion shows that the full-scale solution may be approximated using the the multiscale model shown in Figure \ref{fig:cp}(a).

We partition the problem into two scales, one macroscopic or specimen scale and a second microscopic, representative volume or unit cell scale.  At the macroscopic scale, we have the usual boundary value problem associated with the balance of momenta.  However, we do not have a constitutive model that relates to the deformation gradient history $\{\bar{F}(\tau):  0<\tau<t\}$ to the (Piola-Kirchhoff) stress $\bar{S}(t)$ to close this problem.  Instead, we obtain it from the unit cell problem as follows.  At each point $x$ at the macroscopic scale, at each instant $t$, we extract the deformation gradient history $\{\bar{F}(\tau):  0<\tau<t\}$ of that particular point, and use that as the boundary condition for the unit cell problem.  We solve the crystal plasticity problem over the unit cell subject to this boundary condition, find the complex stress distribution over the unit cell problem at time $t$, average this stress over the unit cell $\bar{S}(t) = \langle S(y,t) \rangle$ where $\langle \cdot \rangle$ denotes spatial average over the unit cell, and return this $\bar{S}(t)$ as the stress at the macroscopic point $x$ at time $t$.  This has also been called the FE$^2$ approach, since one has to solve one finite element problem at the macroscopic scale and a second one at the microscopic scale.  Unfortunately, even this two scale problem is far too expensive to solve at the desired level of accuracy for convergence.  So we seek a surrogate for the unit cell problem as shown in Figure \ref{fig:approach}(c).

\subsection{Recurrent neural operator}


We seek an accurate and easy to train approximation for the map $\Phi: \{\bar{F}(\tau): 0<\tau<t\} \to \bar{S}(t)$.  Note that this map is history dependent, but causal: the stress at time $t$ depends on the entire history of deformation gradient in the time interval $(0,t)$.  We draw inspiration from the established practice in continuum physics by introducing state variables (also known as internal variables or order parameters) that describe the state of the material, and thus carries all the information about the past history that is relevant to the current response.  Thus, these theories are not only causal, but also Markovian (the response only depends on the current state).

We define a recurrent neural operator (RNO)  as a mapping $\Phi_\text{RNO}: \mathcal{F} \to \mathcal{G}$ defined through the relations
\begin{equation} \label{eq:rno}
\begin{cases}
\bar{S}(t) = \psi_S (\bar{F}(t), \xi(t)) \quad & t \in (0,T),\\
\dot{\xi}(t) = \psi_\xi (\bar{F}(t), \xi(t)) \quad & t \in (0,T)
\end{cases}
\end{equation}
where $\psi_S, \psi_\xi$ are feed-forward neural networks, and $\xi$ in a $d_\xi$ dimensional vector of state variables \cite{liu_2023}.    Importantly, in a departure to the practice in continuum physics, we do not postulate or identify the state variable $\xi$ {\it a priori}, but learn it from the data.  

The architecture is causal, i.e, stress only depends on the prior history.  Further, we define it using a ordinary differential equation that is continuous in time.  We can of course discretize it as we see fit.  In the simplest first order, backward Euler, explicit time discretization, we obtain following: in the $(n+1)^\text{th}$ time step, we are given given $\bar{F}^{n+1}, \bar{F}^{n}, \bar{S}^n, \xi^n$, and we find $\bar{S}^{n+1}, \xi^{n+1}$ as
\begin{equation} \label{eq:rnn}
\begin{cases}
{\bar S}^{n+1} = \psi_S (\bar{F}^{n+1}, \xi^n), \\
\xi^{n+1} = \xi^n + \Delta t \ \psi_\xi (\bar{F}^{n+1}, \xi^n) 
\end{cases}
\end{equation}
where $\Delta t$ is the time step.  This is shown in Figure \ref{fig:cp}(b).  

This discrete approximation (\ref{eq:rnn}) has the structure similar to a recurrent neural network (RNN) that has been widely used in natural language processing and other fields.  Indeed RNNs have also been used to model plasticity \cite{ghavamian_2019,mozaffar_2019,wu_2020,bonati_2022}.  This motivates the terminology RNO.  However, there are two critical differences between RNO and RNN.  First, we only have two feed forward neural networks instead of a complex collection of gates and neural networks.  Thus, this architecture is simpler and involves fewer hyper-parameters, and thus requires less data for training.   Second, and crucial for multiscale modeling, the architecture -- the feed forward networks $\psi_S, \psi_\xi$ and the state variables -- are completely independent of the time-step.  Therefore, RNO is discretization independent.  Indeed, it is not only independent of the time step, but we can use any  time discretization scheme (implicit, explicit, Runge-Kutta etc.\ at any order) required by the specific problem.  This is not true for RNNs that are only defined in a discrete setting and thus can only be trained for a particular time discretization scheme and specific choice of time step.  RNNs were developed for natural language processing which has a very different history dependance, and is inherently discrete.  In contrast, we are interested in physics that has a different history dependence and is continuous in time.  RNOs are designed with this in mind.  

Note that the RNO (\ref{eq:rno}) has the structure that is common in continuum physics.  The first of these equations may be regarded as the stress-strain (or any measurable intensive-extensive conjugate variable pair) while the second equation is the evolution equation for the state variable (flow rule of plasticity, time-dependent Landau-Ginzburg equation etc.).  However, unlike traditional models of continuum physics, we do not define the state variables or order variable {\it a priori}, but learn it from the data.  Thus, we may regard RNO as a means of discovering the state variable or order parameter from macroscopic data.

Variations of this architecture are possible.  For example, in the study of viscoelasticity, it is useful to let $\psi_S, \psi_\xi$ depend on $\dot{\bar F}$ in addition to $\bar{F}$ and $\xi$ \cite{bhattacharya_2023}.  Or, we can drive  $\psi_S$ and $\psi_\xi$ from a elastic energy density and dissipation potential respectively, thereby endowing an RNO with a variational or gradient flow interpretation enabling fast and accurate implicit time discretization.

\subsection{Results}

We apply the RNO to build a surrogate model for the unit cell problem of crystal plasticity shown in Figure \ref{fig:cp}(a).  We generate data as shown in Figure \ref{fig:cp}(c) from a polycrystalline unit cell consisting of 1000 equiaxed grains, by subjecting it a set of  input strain histories $\bar{F}(t)$, and computing the corresponding output of averaged stress histories $\bar{S}(t)$.  We use a model for magnesium  that consists of the basal, prismatic and pyramidal slip along with tensile and compressive twinning (treated as pseudoslip) \cite{chang_2015}.  Each component of the input deformation gradient $\bar{F}$ independently changes rates at random times to random values.  A few components are shown in Figure \ref{fig:cp}(c).  The average stress history is computed using Taylor averaging.  The details are provided in Liu {\it et al.} \cite{liu_2022}.

The data consisting of 2400 deformation gradient and stress history pairs $\{\bar{F}^{(n)}(t), \bar{S}^{(n)}(t) \}_{n=1}^{2400}$ is used to train an RNO where the neural networks $\psi_S, \psi_\xi$ have 5 hidden layers with 200 nodes in each layer.  While the state variables are learnt from the data, we need to specify the number (dimension) of state variables.  So, we train the RNO assuming varying number of state variable ranging from zero to 20.  The result is shown in Figure \ref{fig:cp}(d).  We see that the test error is high when we have zero state variables, but drops as we add state variables.  However, the error saturates at five state variables: adding additional state variables does not decrease either the training or test error.  Thus, we conclude that we need five state variables to adequately describe the behavior of our polycrystalline unit cell.  

The figure also shows the error of another neural operator (PCA-Net described later) that is not causal.  It requires significantly larger amount of data to train, but still has larger errors.  Finally, the classical Johnson-Cook model fitted to the training data had an average error of greater than 0.2 in reproducing the test data.  In short, the RNO can be trained with a reasonable amount of data to a high level of accuracy.

Once trained, the RNO is used as a constitutive relation in solving macroscopic boundary value problems.  An example of the Taylor test -- where a cylinder traveling at a uniform velocity slams against a rigid wall leading to waves and inelastic deformation -- is shown in Figures \ref{fig:cp}(e,f).  Figure \ref{fig:cp}(e) shows the impact force as a function of time computed using three distinct time resolutions.  We see that the response is independent of time resolution, demonstrating the time-resolution independence of the RNO (the slight discrepancy near the peak force for the coarsest resolution is due to inadequate resolution of waves in the macroscale solver at the coarsest resolution, instead of the inadequacy of the RNO).  Figure \ref{fig:cp}(f) shows the computational cost of using an RNO in a macroscale solver.  We see that it scales linearly with the size, and is only marginally more expensive than that of a classical Johnson-Cook model.  There is a one-time off-line cost of acquiring the data and training the model, but this is orders of magnitude smaller than a single poorly resolved FE$^2$ calculation.  Finally, we conduct {\it a posteriori} error analysis by extracting deformation gradient trajectories from the macroscale simulation, and comparing the error between the polycrystal calculation and the RNO for these trajectories.  The details as well as various other tests are described in \cite{liu_2023}.

\subsection{Discussion}

The  results demonstrate that an RNO is an excellent surrogate, and provides multiscale or FE$^2$ accuracy at a computational cost comparable to that of a classical constitutive relation.  We have also studied its use in viscoelasticity \cite{bhattacharya_2023}, in architected metamaterials \cite{zhang_2024}, in friction \cite{liu_2025} and in chemically reacting flows through porous media \cite{karimi_2024}.  We discuss a variety of open questions.

We saw above that the behavior of the polycrystal is described by five state variables.  An examination of the response shows that the polycrystal displays yield, isotropic as well as kinematic hardening.  A classical theory of plasticity requires eleven state variables: five for plastic strain, one for isotropic hardening and five for kinematic hardening.  The fact that we are able to represent this data with five is, therefore, surprising.  It may suggest a redundancy in classical models, or a limitation in our dataset.  Classical models are typically developed incrementally, introducing and adding state variables from intuition as we consider progressively complex behavior.   In contrast, the RNO learns the state variable from the data, and all the data at once.  So the RNO is a good way of obtaining an unbiased view of the state variables that describe the entire data.  However, it is difficult to interpret, and a systematic approach to find an intuitive or experimentally measurable interpretation for the state variables learnt by the RNO is an area of active research.

The lower scale model was sampled on a family of trajectories that are independent of any macroscopic simulation.  However, the imposition of balance laws and the nature of constitutive behavior constrain the actual trajectories.  While the sampling approach gave good accuracy in macroscopic simulation as assessed by {\it a posteriori} estimates,we can improve the accuracy by iterative training: first training the model with random trajectories as described above, using a macroscopic calculation to generate new trajectories, transfer learning using a new data set consisting of a combination of the random and simulation trajectories, and iterating \cite{zhang_2024}.  Still, a systematic approach to quantifying physically relevant trajectories and uncertainty remains open.  

We used numerical data from small scales simulations in this work.  However, we could have also used data from experiments.  There are two issues one has to address.  This approach requires significant amount of data, and experimental data is typically sparse.  So one idea is to use multiscale numerical data to train an initial RNO, and then use the sparse experimental data to refine the RNO using transfer learning.

\section{Composite materials and Fourier neural operator}

\subsection{Microstructure and fields}

Consider a composite material where the microstructure scale is small compared to the overall scale of the structure.  For specificity, we consider the problem of diffusion in a periodic medium following \cite{bhattacharya_2024}.   The classical theory of homogenization states that we can find the effective diffusivity of the medium in the direction $\hat e$ by solving an unit cell problem, 
\begin{equation} \label{eq:diff}
 - \nabla \cdot D (x) (\nabla c (x) - \hat {e})  = 0, \quad c \ \text{periodic} 
\end{equation}
where $D$ is the diffusivity tensor (possibly anisotropic), $c$ is the concentration and $\hat{e}$ is the average concentration gradient.  In some situations, we may want to know not only the overall diffusivity but also the full field concentration and its gradient. Therefore, we are interested in learning the map $\Phi: D \to u$.  Note that while the differential equation (\ref{eq:diff}) is linear in $u$, the map $\Phi$ is nonlinear and non-local.  We use a neural operator to approximate this map.

\subsection{Neural operator heuristics}

The key idea of neural operators is to combine the discrete representation of functions with neural networks \cite{bhattacharya_2021,kovachki_2023}.  We use the former in all of computational sciences including computational materials science.  Most models of materials are described by partial differential equations involving functions.  To use digital computers, we represent these functions using discrete vectors of values by introducing a basis set (sinusoidal function in periodic Fourier representation, spherical harmonics in atomic orbitals, or nodal basis functions in finite elements) or an interpolation rule (splines or finite difference).   We choose our basis set to be sufficiently rich so that we can approximate all functions of interest.  In other words, we implicitly introduce two mappings: a mapping $\mathcal P$ that takes functions in some function space $\mathcal F$ and represents them using a d-dimensional vector (in ${\mathbb R}^d$), and a second mapping $\mathcal L$ that takes a d-dimensional vector (in ${\mathbb R}^d$) and identifies it with a function in $\mathcal F$.  We choose these maps such that their composition approximates identity, i.e., if we pick some function in $\mathcal F$, apply $\mathcal P$ to it followed by $\mathcal L$, we recover an approximation to the original function.

\begin{figure}
\centering
\includegraphics[width=5in]{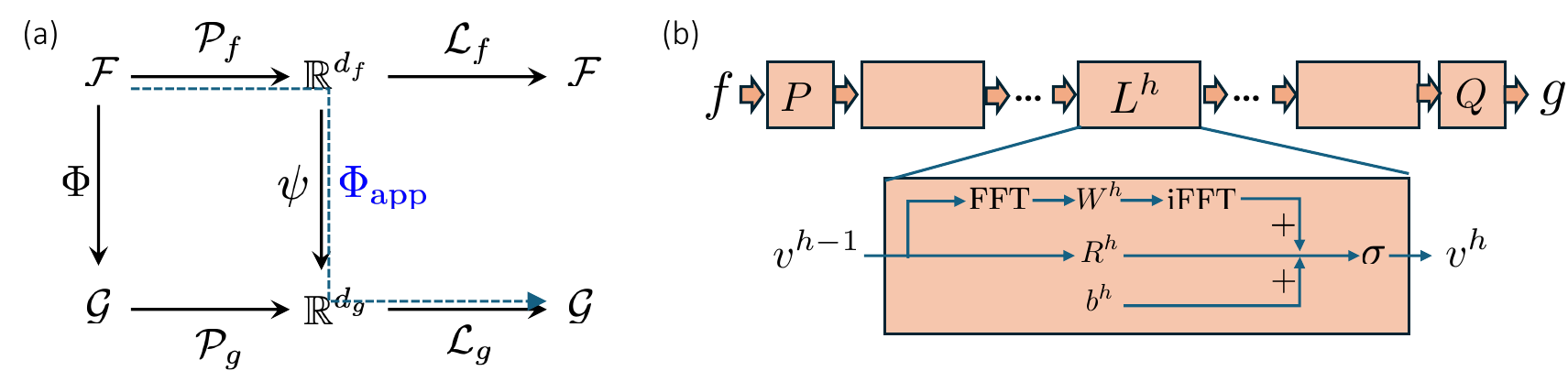}
\caption{\small (a) Basic idea of neural operators.  Adapted from \cite{kovachki}.   (b) Fourier neural operator. \label{fig:no}}
\end{figure}
The basic idea of neural operators is shown in Figure \ref{fig:no}(a).  We seek to approximate an operator $\Phi$ that maps a function in function space $\mathcal F$ to a function in function space $\mathcal G$.  We introduce a  pair of mappings ${\mathcal P}_f$, ${\mathcal L}_f$ that take functions in $\mathcal F$ to a $d_f$ dimensional vector and then back to $\mathcal F$, and the corresponding pair  ${\mathcal P}_g$, ${\mathcal L}_g$ that take functions in $\mathcal G$ to a $d_g$ dimensional vector and back to $\mathcal G$, such that their compositions approximate identity.  We also introduce an appropriate neural network $\psi$ that maps a $d_f$ dimensional vector to a $d_g$ dimensional vector.   We define $\Phi_\text{app}$ to be a composition of ${\mathcal P}_f$ followed by $\psi$ followed by  ${\mathcal P}_l$:
\begin{equation}
\Phi_\text{app} (f) := {\mathcal L}_g ( \psi ( {\mathcal P}_f (f))) \approx \Phi(f) \quad \forall \ f \in K
\end{equation}
where $K$ is the admissible class of inputs.  If we use principal component analysis (PCA) for the model reduction (${\mathcal P}, {\mathcal L}$) and neural networks for the nonlinear map $\psi$, then we obtain a simple architecture called PCA-Net \cite{bhattacharya_2021}.  Many other architectures are possible using the same idea of combining model reduction and nonlinear map, including the Fourier neural operator we describe next.

\subsection{Fourier Neural Operator}

We define a Fourier neural operator (FNO) $\Phi_\text{FNO}: \mathcal F \to \mathcal G$ through the relation,
\begin{equation} \label{eq:fno}
\begin{aligned}
&g = \left(  Q \circ \sigma(v_{H-1}) \circ \cdots  \circ \sigma(v_{h-1}) \circ \cdots  \circ \sigma(v_0)  \circ P \right) f,
\end{aligned}
\end{equation}
that is a composition of an input layer $P$, $H$ hidden, nonlinear, nonlocal layers $\sigma(v_{h-1}), h= 1, \dots H $, and an output layer $Q$ in the language of neural networks \cite{li_2021,kovachki_2023}.  This is shown in Figure \ref{fig:no}(b).  While the implementation is discrete, it is conceived to be an operator on functions, and so we describe this first.  The input is a function on domain $D$,  $f: D \to \mathbb{R}$.  $P$ lifts it linearly to a $d_c$-valued function $v^0: D \to {\mathbb R}^{d_c}$ ($v^0(x) = \{v^0_1 (x) \dots v^0_{d_c} (x) \}$) where each $v^0_p$ is linear in $f$.  Then, we have a series of nonlinear Fourier layers that maps the functions $v^{h-1}: D \to {\mathbb R}^{d_c}$ to $v^{h}: D \to {\mathbb R}^{d_c}$ as follows,
\begin{equation}
v_p^h(x)  = \sigma \left( \sum_q (W^h_{pq} v_p^{h-1} (x)) + \left({\mathcal F}^{-1} \left( \sum_{q} R^h_{pq}(k) (\mathcal{F} v_q^{h-1})(k) \right)  \right)(x) + b^h_p \right) 
\end{equation}
where $\sigma$ is a nonlinear activation function,  $W^h$ is a $d_c \times d_c$ matrix valued function of $x$,  $R^h$ is a $d_c \times d_c$ complex matrix valued function of Fourier variable $k$, and $\mathcal F$ and $(\mathcal F)^{-1}$ are the Fourier and inverse Fourier transform respectively.  Finally $Q$ linearly projects the $d_c$ valued function $v_H: D \to {\mathbb R}^{d_c}$ to the output $g: D \to \mathbb R$.

Now, if the input is discrete with a discretization $N$, $i$ is an $N$ dimensional column vector while $v^h$ is a $N \times d_c$ dimensional matrix.  $P,Q$ are $1 \times d_c $ row vectors (so $v^0_{pq} = i_p P_q$, $o_p = v^H_{pq} Q_q$). The Fourier transform is now discrete and taken column by column on $v^h$.  We limit it to $M$ modes.  Thus $R$ is $M\times d_c \times d_c$ complex valued matrix.  Thus,
\begin{equation}
v_p^h(x_n)  = g \left( \sum_j (W^h_{pq} v_q^{h-1} (x_n)) + \left({\mathcal F}^{-1} \left( \sum_{j} R^h_{mpq} (\mathcal{F} v_q^{h-1})(k_m) \right)  \right)(x_n) + b^h_p \right)
\end{equation}
for $n = 1, \dots, N$.  The unknown parameters are $\{P, Q, W^h, R^h, b^h\}$, and these are independent of discretization $N$.  $H$ is called the depth of the network and $d_c$ is called the width or number of channels.

An open source implementation of FNO is available through GitHub \cite{kossaifi_2025}.

\subsection{Results}
\begin{figure}[t]
\centering
\includegraphics[width=6in]{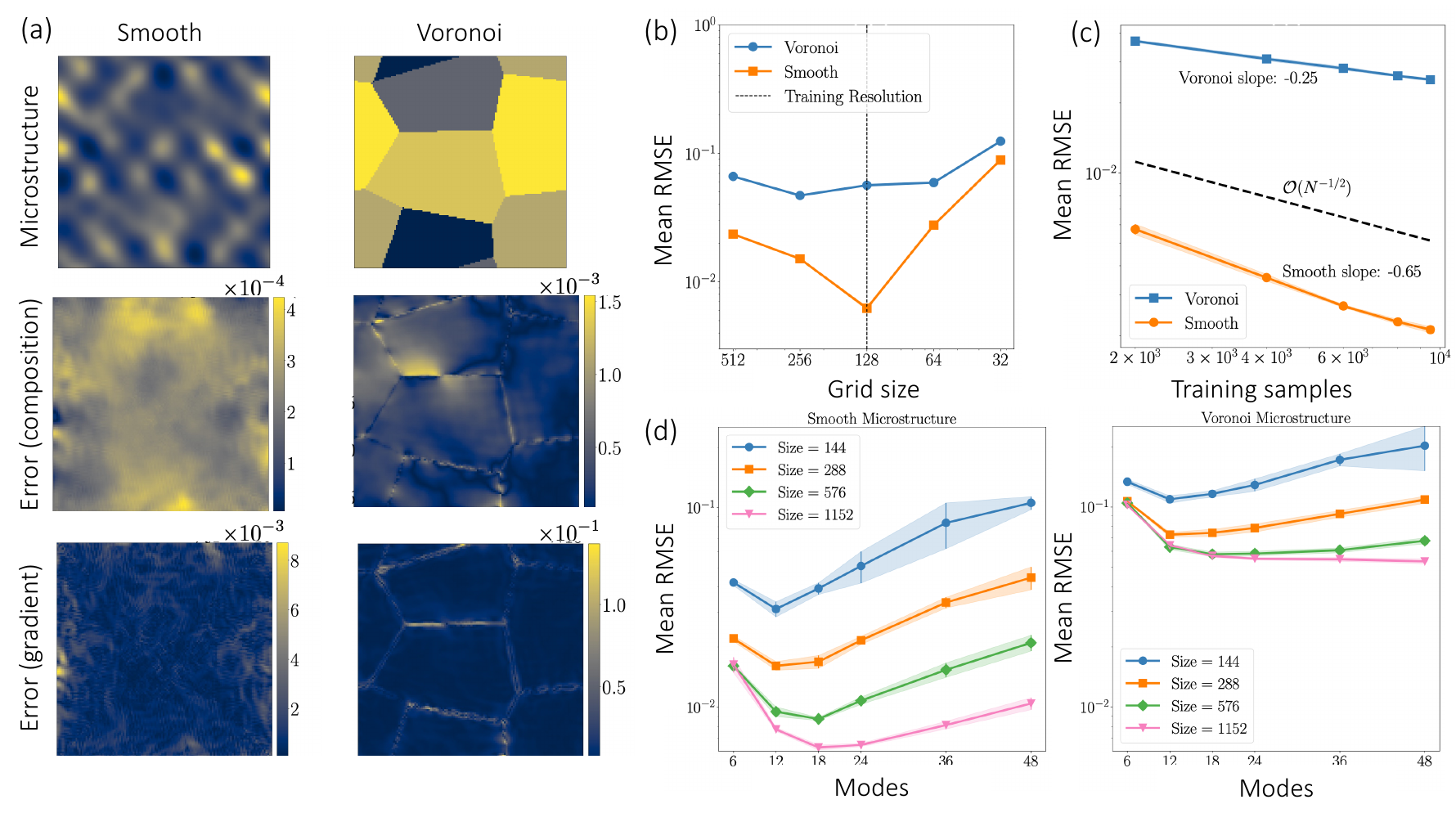}\\ \vspace{0.1in}
(e)
\begin{tabular}{|c|c|c|c|c|c|}
\hline
\small Structure & \small Mean RMSE & \small Median RMSE  & \small Mean RME & \small Median RME & \small Median RDE\\
\hline
\small Smooth &  \small 0.0053 & \small 0.0051 & \small 0.0074 & \small 0.0067 & \small 0.00058\\
\hline
\small Voronoi & \small 0.0501 & \small 0.0450 & \small 0.1864 & \small 0.1686 & \small 0.00155\\
\hline
\end{tabular}
\caption{\small Composite materials.  (a) The point-wise square error in the learnt potential and its gradient (error between the actual solution and the neural operator inference) for one example each drawn from the test sets for the two classes of microstructure.  (b) Resolution independence of error: the neural operator is trained with data at a grid size of $128^2$, but evaluated on data at various resolution.  (c) Scaling of the relative mean error with the size of the training data and (d) Relative error vs. the number of Fourier modes.  (e) Errors as defined in (\ref{eq:errors}).   Adapted from \cite{bhattacharya_2024}.  \label{fig:comp}}
\end{figure}

We now return to diffusion in (\ref{eq:diff}), and the problem of finding a surrogate for the operator from the microstructure $D(x)$ to the concentration $u(x)$.   We consider two classes of microstructures  in two dimensions as shown in Figure \ref{fig:comp}(a):
\begin{itemize}
\item Smooth: $D$ is smooth with the diagonal elements generated using truncated, rescaled log-normal random fields. 
\item Voronoi: $D$ is isotropic but piecewise constant taking a distinct value in each domain, with the domains generated by Voronoi tesselation starting with a random seed.  
\end{itemize}
We generate 9500 data samples by solving (\ref{eq:diff}) using finite elements with a very fine mesh, and sampling the solution on a 128 $\times$ 128 grid.  We train an FNO for 300 epochs with the same architecture of 24 Fourier modes, a model width of 48, and 4 layers using the following loss function.
\begin{equation}
    \text{Loss}= \frac{1}{N}\sum_{n = 1}^N \Bigl(\|c^{(n)} - \widehat{c}^{(n)}\|^2_{L^2} + \|\nabla c^{(n)} - \nabla \widehat{c}^{(n)}\|^2_{L^2}\Bigr)
\end{equation}
where $c$ is the true solution and $\widehat{c}$ is the FNO approximation.  We test the results on 500 data samples generated similarly and independent of the training sample.  We use three measures of error:
\begin{subequations}\label{eq:errors}
    \begin{equation}
    \text{Relative root-mean-square (RMSE)} = \frac{1}{N}\sum_{n=1}^N \left(\frac{\|c^{(n)}- \widehat{c}^{(n)}\|^2_{L^2} + \|\nabla c^{(n)} - \nabla \widehat{c}^{(n)}\|^2_{L^2}}{\|c^{(n)}\|^2_{L^2} + \|\nabla c^{(n)}\|^2_{L^2}}\right)^{\frac{1}{2}} 
    \end{equation}
    \begin{equation}
    \text{Relative maximum  (RME)} = \frac{1}{N}\sum_{n=1}^N \left(\frac{\|c^{(n)}- \widehat{c}^{(n)}\|^{10}_{L^{10}} + \|\nabla c^{(n)} - \nabla \widehat{c}^{(n)}\|^{10}_{L^{10}}}{\|c^{(n)}\|^{10}_{L^{10}} + \|\nabla c^{(n)}\|^{10}_{L^{10}}}\right)^{\frac{1}{10}} 
    \end{equation}
\begin{equation}
    \text{Relative $\overline{D}$ (RDE) } = \frac{\|\overline{D} - \widehat{\overline{D}}\|_F}{D_u - D_l} 
\end{equation}
\end{subequations}
The first is the usual relative root-mean-square error combining the solution and its gradient.  The second is the error in the tenth power of the solution and this serves as a proxy for the maximum error in the pointwise values in the solution and its gradient.  Finally, we can use the solution to compute the effective diffusivity $\overline{D}$, and the third error provides the relative error in the effective diffusivity.  Note that we normalize it using the difference between the Voigt  upper bound $D_u$ and Ruess lower bound $D_l$ rather than $D$ since this gives a more sensitive view of the error.

The results are shown in Figure \ref{fig:comp}.  Figure \ref{fig:comp}(a) shows the result from a typical test sample each of the smooth and Voronoi microstructures.  We see that the error in the composition and its gradient is small but largely uniformly distributed across the specimen in the case of the smooth microstructure; in contrast, the errors are concentrated at the grain boundaries and larger in the case of the Voronoi microstructure.  This is made precise in the table in Figure \ref{fig:comp}(e) using the mean and median value of error across all 500 test specimens.  We find that the mean and median are similar.  Further, we find that the RMSE and RME are similar in the smooth case, while they are quite different in the Voronoi case as anticipated from Figure \ref{fig:comp}(a).  We also find that the errors are larger for the Voronoi case compared to the smooth case.  However, we find that the errors in the effective conductivity are small for both cases.

Figure \ref{fig:comp}(b) demonstrates the discretization independence of the FNO architecture.  We train the FNO on data samples on a a $128 \times 128$ grid as described, but test it against data generated at various grid sizes ranging from  $32 \times 32$ grid to a $512 \times 512$ grid.  We see very little variation.  A standard neural network would have the desired accuracy at the training grid size, but suffer uncontrolled errors at other grid sizes \cite{bhattacharya_2021}.

Figure \ref{fig:comp}(c) shows the data required for training the FNO.  We see that the smooth microstructures are more amenable to training compared to the Voronoi.  Figure \ref{fig:comp}(d) shows the interplay between the model width and Fourier modes in an FNO.  It shows how the error as a function of the number of Fourier modes and the model size (model width $\times$ Fourier modes).  Clearly increasing the model size reduces the error in both cases.  However, increasing the Fourier modes while keeping the model size initially decreases the error, but eventually increases it.  This shows that one needs a good balance between the model width and Fourier modes.

\subsection{Discussion}

The results above demonstrate that an FNO can learn the nonlinear map from the microstructure to the concentration in a composite medium.  We find excellent agreement for smooth variations in the diffusivity (even with large contrast), but poorer -- though still good --  agreement with discontinuous diffusivity.  This is because of the inability of the Fourier representation to adequately represent the discontinuity in concentration gradients at the interfaces.  A potential idea to improve the accuracy is to enhance the FNO with a local kernel  \cite{liuschiaffni_2024}.

The FNO has been widely used in a variety of problems, including Navier Stokes equation \cite{kovachki_2023},  elastic wave propagation \cite{lehman_2024}, ocean modeling \cite{choi_2024} and quantum spin dynamics \cite{shah_2026}.
The Fourier neural operator is one of a number of different possible manifestations of a neural operator \cite{kovachki_2023}.  The PCA-net described earlier is the simplest.  The graph neural operator, combined with ideas of multipole expansion, uses a graph  to represent spatial interpolation \cite{li_multipole_2020}.

\section{Density functional theory}

\begin{figure}
\centering
\includegraphics[width=6in]{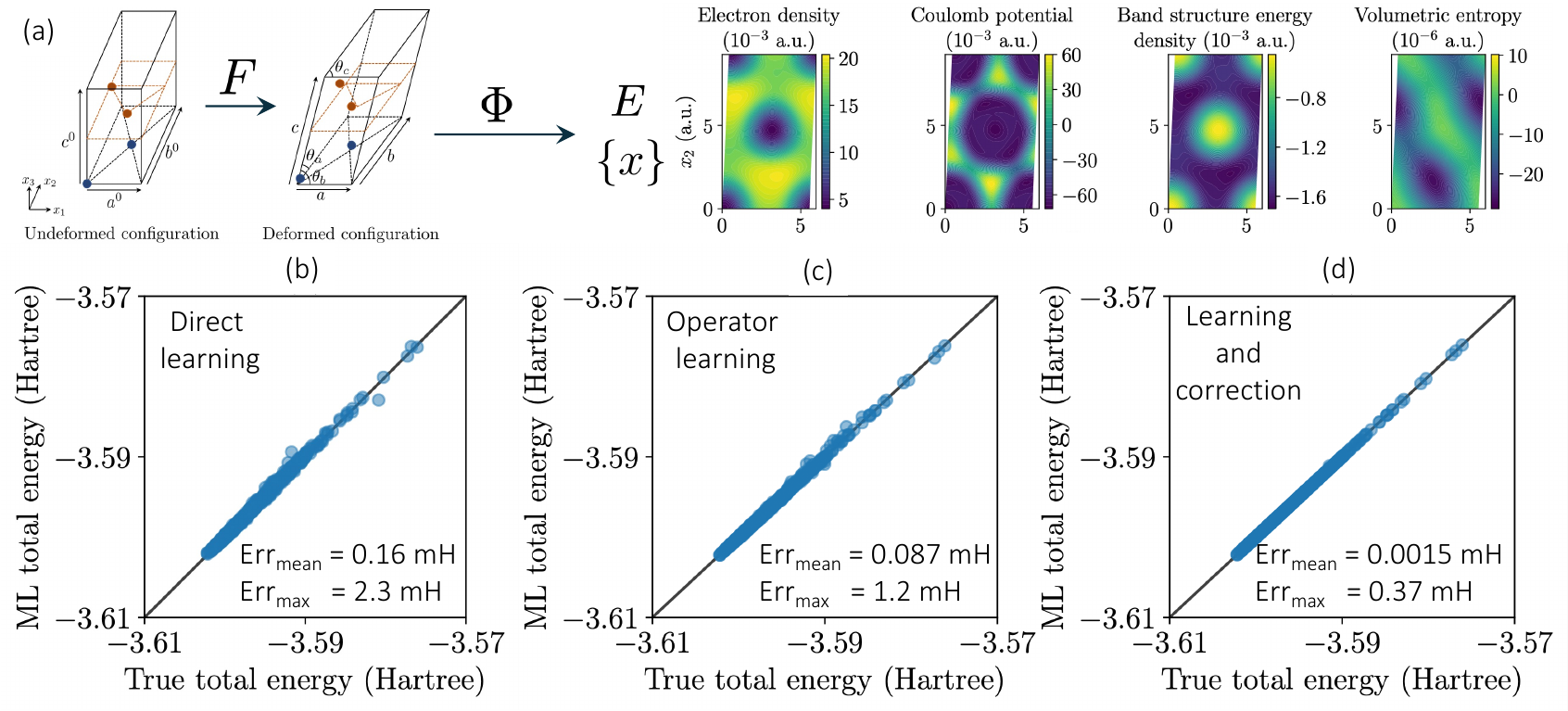}
\caption{\small Density functional theory.  Adapted from \cite{teh_2021} \label{fig:dft}}
\end{figure}

Neural networks and neural operators are trained using stochastic gradient descent, and thus susceptible to getting stuck in local minima.  Further, while universal approximation theorems provide existence, it does not provide a design.  Finally, we sample the data.  For all these reasons, we expect accuracy in a probability (or mean) sense.  But there are problems in materials science where one needs to guarantee maximum error.  One such case is the use of first principles density functional theory (DFT) to study stability of structures by computing and comparing the energy of competing structures.  This requires us to  compute small differences between two large numbers, and therefore it is very important to guarantee good accuracy.  The final example following \cite{teh_2021} shows how a neural operator surrogate can be used to learn additional information to improve on the accuracy of a neural network surrogate.  

We consider the problem of understanding the energy and stability of the hexagonal close-packed structure as we deform it.  We describe the deformation relative to the ground state using a $3\times 3$ (deformation gradient) matrix $F$ (see Figure \ref{fig:dft}(a)), and seek to learn the total energy $E$ as a function of $F$, $\psi: F \to E$. We generate data  data $\{F, E(F)\}$ by sampling $F$, and solving the equations of DFT.  We use a spectral quadrature approach \cite{ghosh_2017a,ghosh_2017b} to solve the equations of DFT, though we could use other solvers.  We use this data to train a neural network (since both $F$ and $E$ are finite dimensional) to learn the map $\psi$.  The results are shown in Figure \ref{fig:dft}(b); the root mean square test error is 0.16 mHartree which is well within the desired accuracy of 1.5 mHartree.  However, the maximum test error is above this value, and the neural network surrogate is thus not entirely reliable. 

To improve upon it, we observe that DFT actually solves a nonlinear problem to obtain the electronic structure in the form of fields or spatially dependent functions, and then uses the electronic structure to compute the total energy.  In the spectral quadrature approach, we compute four fields $u_i(x), i=1,\dots,4$ (electronic density, Coulomb potential, band structure energy and electronic entropy; example in Figure \ref{fig:dft}(a)) as well as the equilibrium position of the internal atoms $\bar x$, and then use these to compute the energy: In other words, $\Phi_\text{DFT} = F \to (\{u_i\}, \bar{x}), \ E = E(\{u_i\}, \bar{x})$.  So, we generate the data $(F, (\{u_i\}, \bar{x}))$ in our previous calculation.  But $u_i$ are fields or functions, and therefore we use a neural operator to learn a surrogate for the map $\Phi_\text{DFT}$, and the use these fields to evaluate $E$.  The results are shown in Figure \ref{fig:dft}(c).  We see that the mean and maximum test error has improved dramatically.  The fact that the neural operator learnt the fields that contain significantly more information leads to better learning.

We can gain further accuracy by using the surrogate in the underlying numerical method.  We solve the nonlinear DFT equations by (fixed point or self-consistent field or SCF) iteration.  We typically conduct many tens of iterations, and relaxing the positions of the internal atoms increases this by another tens of factors.  However, we do not need many iterations if we have a good initial guess.  So, we use the output of the surrogate as the initial guess, and then use one SCF iteration.  The results are shown in Figure \ref{fig:dft}(d): we get remarkable accuracy.  Indeed, the maximum error of 0.37 mHartree is comparable to the inherent accuracy of DFT itself.

\section{Discussion and outlook}

This short article describes three examples of the use of neural operators in multiscale modeling.  In the first, we use a recurrent neural operator (RNO)  in the temporal domain to learn the history dependance of crystal plasticity in a polycrystal for use in macroscopic simulations.  In the second, we use a Fourier neural operator (FNO) in the spatial domain to learn the microstructure to solution map in a composite medium.  The third shows how a neural operator can used even in situations where it is necessary to guarantee high accuracy.

Multiscale modeling and neural operators are a very natural combination.  Multiscale modeling is essential in understanding the properties of materials.  Till recently, the transfer of relevant, and all relevant, information from one scale to another remained a challenge in multiscale modeling.  Neural operators, trained on data generated by the high fidelity solution of the fine scale, and used as surrogates at the larger scales, provide an attractive avenue to overcoming this challenge.   Conversely, multiscale modeling is an ideal application of neural operators.   While they are inexpensive to use once they are trained, there is a cost associated with the generation of data and with training.  Multiscale modeling requires the repeated use of the fine scale problem, and thus one can amortize the cost of generating the data and training.  Further, neural operators only guarantee accuracy in some statistical (root-mean-square) sense.   This is also typically acceptable in multiscale modeling where we only require accuracy in some integral norm.

This use of a neural operator (or networks) as a surrogate is distinct from their use to solve a partial differential equation, or to represent a partial differential equation in data assimilation.  There is a large literature on this going back to the physics-informed neural networks (PINNs) \cite{raissi_2018}.  Neural operators, and in particular FNOs, have also been used in this setting \cite{li_2024}.  We do not discuss this application of neural operators here.

This article focussed on the computational aspects of the use of neural operators in multiscale modeling.  However, there is another exciting prospect, one of discovering new physics, that is the subject of current research.   We discuss two such directions.  First, recall that the RNO was able to learn the state variables from the stress and strain histories, and these state variables were distinct from those used in classical theories.  In the latter, we postulate the state variables based either on intuition or the need to augment another empirical theory, but typically lack either a systematic derivation or full understanding.  The discovery of state variables provides new insight into the phenomenon.  However, a full articulation of this insight requires the development of methods to interpret the learned state variables.  Second, in the examples above, we assumed scale separation and used computational data.  Instead, we could use experimental data without the need for scale separation.  Thus, one could learn not only the coarse scale behavior but also the physics of the fine scale.  However, this requires a sufficiently rich class of experimental data.  This goes beyond automating current experimental techniques to obtain high throughput, and requires the development of new experimental techniques that explore the entire phenomenon (for example arbitrary strain paths in plasticity).

There are other related directions to build on the reported work.  First, we separately used RNOs for temporal problems and FNOs for problems in the spatial domains.  However, there are a number of problems in the spatio-temporal domain.  One can combine the two, for example treating the RNO (\ref{eq:rno}) as differential equations on function spaces, and then using FNO (\ref{eq:fno}) as a method to approximate the forward map.  Second, the examples were limited to two scales.  Second, the approach can readily be extended to multiple scales by building a cascade of surrogates.  Finally, we use an intuitive but still {\it ad hoc} approach to sample the trajectories (RNO) and microstructures (FNO) to generate our data.  This is because a systematic theory for characterizing distributions that arise in applications is still lacking.  This can potentially be overcome by using generative neural networks to generate samples based on experimental observations.    

\paragraph{Competing interests} None
\paragraph{Funding} Army Research Laboratory, the Army Research Office, the Office of Naval Research and the De Logi foundation.
\paragraph{Authors' contributions}  Not applicable, sole author
\paragraph{Acknowledgement}
This article is based on a years long collaboration with Andrew Stuart and Anima Anandkumar, and reflects the contributions of numerous graduate students and postdoctoral scholars.  It specifically draws on the work of Kamyar Azzizadenasheli, Swarnava Ghosh, Nikola Kovachki, Zongyi Li, Burigede Liu, Eric Ocegueda, Ying Shi Teh and Margaret Trautner.


\end{document}